# Angular Distribution of Bremsstrahlung Produced by 10-keV and 20-keV Electrons Incident on a Thick Au Target

D. Gonzales and S. Williams

*Department of Physics, Angelo State University, San Angelo, Texas 76909, USA*

**Abstract.** The relative intensities of the thick-target bremsstrahlung produced by 10-keV and 20-keV electrons incident on Au at forward angles ranging from 0° to 25° are compared. Following corrections for photon absorption within the target, the detected radiation appears to be distributed anisotropically only for photon energies, $k$, that are approximately equal to the initial energy of the incident electrons, $E_o$. The results are compared to the theoretical angular distributions of Kissel *et al.* [At. Data Nucl. Data Tables **28**, 381 (1983)]. The comparison suggests that when $k/E_o \approx 1$, the angular distribution of bremsstrahlung emitted by electrons incident on thick targets is similar to the theoretical angular distribution of bremsstrahlung emitted by electrons incident on free-atom targets.

## INTRODUCTION

In previous experiments we have studied the angular dependence of thick-target bremsstrahlung (TTB) produced by electrons with initial energies in the range of 10 to 20 keV incident on Ag [1]. TTB differs from so-called "thin-target" bremsstrahlung in that the target is thick enough that each incident electron is likely to interact with several target atoms (emitting multiple photons in the process), rather than just a single one. After correcting for several factors including photon attenuation due to the target itself, the data suggested that as photon energies, $k$, approached the initial electron energies, $E_o$, (i.e. $k/E_o \rightarrow 1$) the angular distribution of the emitted radiation was anisotropic. Conversely, as $k/E_o \rightarrow 0$, the emitted radiation was distributed isotropically. The data were also compared to the theoretical bremsstrahlung angular distributions ("shape functions") for Ag as tabulated by Kissel *et al.* (KQP) [2]. The KQP shape functions were calculated for situations where electrons are incident on free atoms (rather than solid, thick targets). The anisotropic angular distribution observed in the experimental TTB data for situations where $k/E_o \approx 1$ was shown to be similar to the theoretical angular distribution of bremsstrahlung emitted by electrons incident on free-atom targets. This behavior was attributed to the fact that photons with energies that are approximately equal to the energies of the incident electrons are likely emitted before the initial direction of the electrons has been changed through scattering. Thus, photons with energies $k \approx E_o$ are typically emitted by electrons that have interacted with only a single target-atom (near the surface of the target) and one would expect the angular distribution of the emitted bremsstrahlung to be similar to the free-atom target KQP shape functions. However, the angular distribution of the *detected* bremsstrahlung will be affected by photon scattering inside the target and attenuation due to the target itself; thus, attenuation due to absorption in the target must be accounted for in order for this trend to be observed. Furthermore, when $k/E_o < 1$ multiple collisions within the target may have occurred before the photon was emitted. Since scattering leads to electrons' directions being randomized, one would expect that as $k/E_o \rightarrow 0$ the emitted radiation would be essentially isotropic. As there have been relatively few experimental studies of the angular distribution of bremsstrahlung [1, 3-5], and none where attenuation due to the target has been considered (other than [1]), there is certainly a need for additional work in the area. The primary goal of this study was to confirm the results of the previous study [1] using 1.932 mg/cm$^2$-thick Au (which, like Ag, was one of the "benchmark" materials of the KQP tabulations) as a target and to improve the uncertainties associated with our data.

## EXPERIMENTAL PROCEDURE

Bremsstrahlung was produced using an Amptek Mini-X X-ray tube mounted on a rotatable stage in such a way that the 1.932 mg/cm$^2$-thick Au target and 23.47 mg/cm$^2$-thick Be end-window were situated above the center. Inside of the Mini-X, electrons were accelerated using voltages of 10 and 20 kV towards the Au target. The resultant X-rays were detected using an LD Didactic GmbH Si-PIN detector with a relatively small (~ 0.5 mm$^2$) window, positioned at a fixed distance from the center of the rotatable stage. The intensities of the X-rays were measured at forward angles ranging from 0° to 25°. The design of the Mini-X prevented us from measuring the bremsstrahlung intensities at angles greater than 25°. As we planned on comparing the ratios of the bremsstrahlung intensities at different angles, factors such as the efficiency of the detector, the solid angle subtended by the detector, total charge incident on the target, and electron backscattering did not need to be considered (each of which were the same in every experiment).

## RESULTS AND DISCUSSION

### Angular Distribution of Bremsstrahlung Produced by 10-keV Electrons Incident on Thick Au

Fig. 1 compares the ratios of the bremsstrahlung intensities produced by 10-keV electrons at 25° and 0° for $k/E_o$ values of 0.6, 0.7, 0.8, 0.9, 0.95, and 1 (KQP note that tabulations for $k/E_o = 1$ are meant to apply to photons with energies "50 eV or so from the exact tip of the spectrum" [2]) . Error was estimated by combining statistical error, the uncertainty in total charge incident on the target, and uncertainty associated with the subtraction of pile-up in quadrature. Attenuation due to absorption within the target was taken into account (for all data presented here) by applying an exponential term,

$$f(E_o, k, \theta) = e^{\left(\frac{\mu_{Au}x_1 + \mu_{Be}x_2}{cos\theta}\right)}, \quad (1)$$

where $\mu_{Au}$ and $\mu_{Be}$ are the photon attenuation coefficients for photons with energy $k$ for Au and Be, respectively (taken from the NIST XCOM database [6]), and $x_1/cos\theta$ and $x_2/cos\theta$ are the distances that a photon travels through the Au target and Be window, respectively. As discussed in our previous report [1], this distance was approximated by assuming that on-average each electron's energy at the continuous-slowing-down approximation (CSDA) range (taken from the NIST ESTAR program [7]) is equal to zero. Additionally, it is assumed that an electron's energy-loss is essentially linear as it travels through the target. Therefore, using the CSDA range, the point at which an electron's energy is equal to $k$ can be predicted. Calculations performed by Pratt et al. [8] suggest that at low energies (i.e. those involved in these experiments) the probabilities of an electron emitting a photon of any energy between its $E_o$ and 0 keV are approximately equal. Thus the average point at which a photon of energy $k$ is emitted is approximated as the midpoint between the target surface and the point where the CSDA range predicts the electron's energy to be equal to $k$.

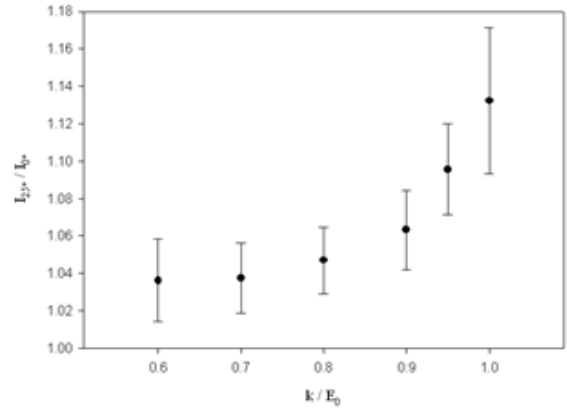

**FIGURE 1.** Comparison of the bremsstrahlung intensity at a forward angle of 25° to the intensity at 0° for various values of $k/E_o$ where $E_o = 10$ keV.

Fig. 1 confirms the results of our previous report [1]. The figure suggests that the anisotropy increases dramatically as $k/E_o \rightarrow 1$. However, it also suggests that for $k/E_o$ values of 0.6 and 0.7, electron-scattering inside the target results in the intensities of the bremsstrahlung at 0° and 25° being essentially equal.

Fig. 2 is a comparison of data obtained from experiments involving 10-keV electrons incident on thick Au to the theoretical free-atom target shape functions of KQP for $k/E_o = 0.95$. The plots are of the ratios of the intensities of bremsstrahlung at an angle θ to the intensities at 0° at angles ranging from 0° to 25°. Although the KQP tabulations predict a much stronger anisotropy, the general trends in the angular distributions are in agreement. This suggests that even when $k/E_o = 0.95$, electron-scattering in the target may affect the angular distribution (experimentally, a 9.5 keV photon could be emitted by an electron that has previously been scattered). Furthermore, photon scattering inside the target may also lead to weaker anisotropy than is predicted for situations involving free-atom targets. KQP predict that the intensity of the emitted bremsstrahlung increases from 0° to approximately 70°, at which point the intensity begins to decrease. It should be noted that KQP conservatively estimate the uncertainties in the shape function calculations to be 5% (and, as they note, less for benchmark materials such as Au). With no exact uncertainties available for the KQP shape functions, we assigned uncertainties of 5%. Therefore the error bars in Fig. 2 represent a very conservative estimate of the error in the theoretical KQP tabulations.

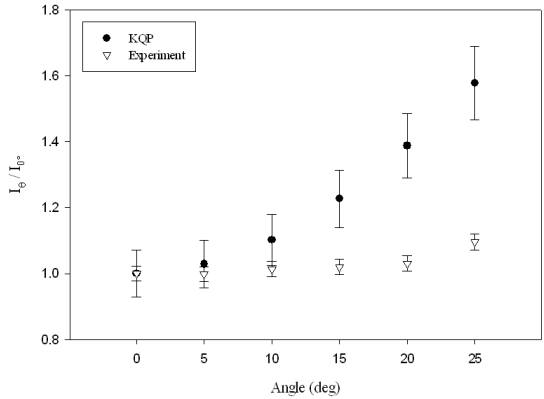

**FIGURE 2.** Comparison of data obtained from experiments involving 10-keV electrons incident on thick Au to the theoretical free-atom shape functions of KQP for $k/E_o$ = 0.95.

Fig. 3 compares the angular distributions of bremsstrahlung for $k/E_o$ values of 0.9 and 0.95. Just as in Fig. 2, the plots are of the ratios of the intensities of bremsstrahlung at an angle θ to the intensities at 0°. As one might expect, the $k/E_o$ = 0.95 data exhibits slightly stronger anisotropy than the $k/E_o$ = 0.9 data. Quantitatively, the ratio $I_{25°}/I_{0°}$ is approximately 1.10 for $k/E_o$ = 0.95 and 1.06 for $k/E_o$ = 0.9. Our data (not shown in this plot) indicates that this trend continues as $k/E_o \rightarrow 0$.

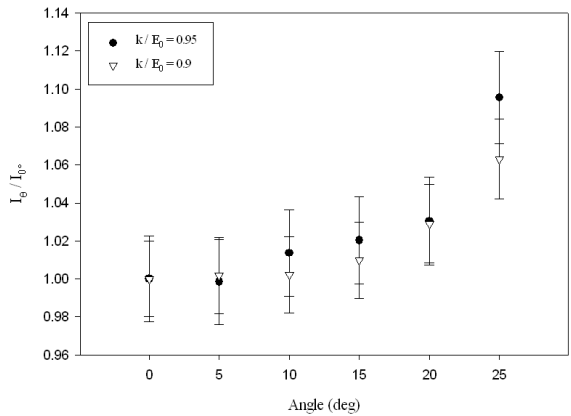

**FIGURE 3.** Comparison of the angular distributions of bremsstrahlung produced by 10-keV electrons incident on thick Au for $k/E_o$ values of 0.9 and 0.95.

## Angular Distribution of Bremsstrahlung Produced by 20-keV Electrons Incident on Thick Au

Fig. 4 is a comparison of the angular distribution of bremsstrahlung produced by 20-keV electrons for $k/E_o$ = 0.9 and 0.6. As in Figs. 2 and 3, the plots are of the ratios of the intensities of bremsstrahlung at an angle θ to the intensities at 0° at angles ranging from 0° to 25°. Data for $k/E_o$ = 0.95 and 1 are not included due to the poor statistical nature of the data and resulting high uncertainties. These plots are similar to Fig. 3; however, as one might expect, due to a larger difference in $k/E_o$ values, the difference in angular distributions is more dramatic. The $k/E_o$ = 0.9 data shows a difference of 16% between the bremsstrahlung intensities at 0˚ and 25˚, while the $k/E_o$ = 0.6 data shows an increase of only 4%. Once again, this suggests that the angular distribution of TTB is similar to that of thin-target bremsstrahlung only when $k/E_o \approx 1$. Otherwise, TTB is essentially distributed isotropically.

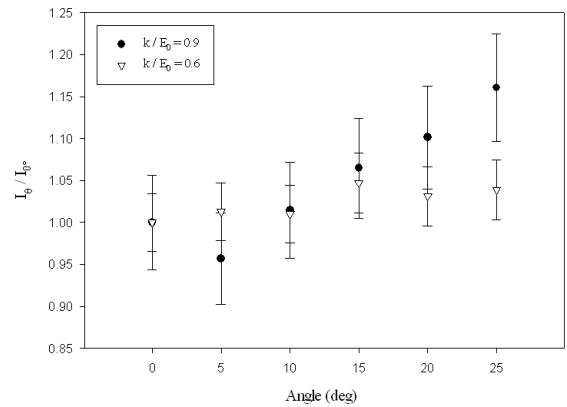

**FIGURE 4.** Comparison of the angular distributions of bremsstrahlung produced by 20-keV electrons incident on thick Au for $k/E_o$ values of 0.6 and 0.9.

## CONCLUSIONS

This report presents the results of experiments in which the angular distributions of bremsstrahlung produced by 10 and 20-keV electrons incident on thick Au were studied. Our results confirm a previous report [1], which suggests that the angular distribution of TTB shares characteristics with that of free-atom target bremsstrahlung only when the emitted photon's energy is approximately equal to the initial energy of the incident electron (i.e. when $k/E_o \approx 1$). As $k/E_o \rightarrow 0$, the angular distribution of the TTB becomes essentially isotropic (once attenuation due to the target is considered), due, primarily, to electron-scattering in the target.

Several experiments are still needed in order to better understand and model the angular distribution of bremsstrahlung. Additional experiments such as those performed by Aydinol et al. [4], involving gas targets, need to be performed in order to confirm the theoretical predictions of KQP. Furthermore, the angular distribution of polarizational bremsstrahlung (which may [9] or may not [10] be present in solid-target experiments) should also be explored.

## ACKNOWLEDGMENTS

This study was partially funded through an Angelo State University Research Enhancement grant. Also, some of the equipment used in this study was obtained through the Department of Energy's Laboratory Equipment Donation Program.